\documentclass[10pt,conference]{IEEEtran}
\IEEEoverridecommandlockouts
\usepackage{cite}
\usepackage{amsmath,amssymb,amsfonts}
\usepackage{algorithmic}
\usepackage{graphicx}
\usepackage{textcomp}
\usepackage{xcolor}
\usepackage{stfloats}
\usepackage{float}
\usepackage[bookmarks=false]{hyperref}
\def\BibTeX{{\rm B\kern-.05em{\sc i\kern-.025em b}\kern-.08em
		T\kern-.1667em\lower.7ex\hbox{E}\kern-.125emX}}
\begin{document}
	
\title{Prema: A Tool for Precise Requirements Editing, Modeling and Analysis \\
}

\author{%
\IEEEauthorblockN{
Yihao Huang\IEEEauthorrefmark{1}, 
Jincao Feng\IEEEauthorrefmark{1}, 
Hanyue Zheng\IEEEauthorrefmark{1}, 
Jiayi Zhu\IEEEauthorrefmark{1}, 
Shang Wang\IEEEauthorrefmark{1}, \\
Siyuan Jiang\IEEEauthorrefmark{4}, 
Weikai Miao\IEEEauthorrefmark{1}\IEEEauthorrefmark{2}\IEEEauthorrefmark{5}\thanks{\IEEEauthorrefmark{5}Geguang Pu, Weikai Miao are the corresponding authors.}, 
Geguang Pu\IEEEauthorrefmark{1}\IEEEauthorrefmark{3}\IEEEauthorrefmark{5} 
}

\IEEEauthorblockA{\IEEEauthorrefmark{1} Shanghai Key Lab of Trustworthy Computing, East China Normal University, China} 
\IEEEauthorblockA{\IEEEauthorrefmark{2} Shanghai Institute of Intelligent Science and Technology, Tongji University, China}
\IEEEauthorblockA{\IEEEauthorrefmark{3} Shanghai Trusted Industrial Control Platform Co., Ltd, China}
\IEEEauthorblockA{\IEEEauthorrefmark{4} Eastern Michigan University, USA}
}%

\maketitle
\begin{abstract}
	We present \emph{Prema}, a tool for \textbf{\textit{P}}recise \textbf{\textit{R}}equirement \textbf{\textit{E}}diting, \textbf{\textit{M}}odeling and \textbf{\textit{A}}nalysis. It can be used in various fields for describing precise requirements using formal notations and performing rigorous analysis. By parsing the requirements written in formal modeling language, \emph{Prema} is able to get a model which aptly depicts the requirements. It also provides different rigorous verification and validation techniques to check whether the requirements meet users' expectation and find potential errors. We show that our tool can provide a unified environment for writing and verifying requirements without using tools that are not well inter-related. For experimental demonstration, we use the requirements of the automatic train protection (ATP) system of CASCO signal co. LTD., the largest railway signal control system manufacturer of China. The code of the tool cannot be released here because the project is commercially confidential. However, a demonstration video of the tool is available at https://youtu.be/BX0yv8pRMWs.
\end{abstract}


\section{INTRODUCTION}
\par Requirements verification and validation (V\&V) is an important research area in requirement engineering~\cite{bibitem13}. They can reduce the number of defects before software deployment~\cite{bibitem12}. V\&V is especially important for \emph{safety-critical systems} which tight compliance with certification processes~\cite{bibitem4}. 
\par Specifications in the industry are commonly specified in natural language, due to the ambiguity of which, rigorous V\&V techniques cannot be utilized. On the other hand, tools designed for requirement engineering are scarce. Doors~\cite{bibitem11} is one of the requirement management tools which is widely used. However, the large number of built-in features of the tool are designed for general purpose. That is, for people in a particular field, who only need limited features of the tool, other unused features are redundant. Since V\&V requires precise specifications, for example, mathematically defined, formal specifications are preferred. To help requirement engineers create and analyze formal specifications, tools~\cite{bibitem12} for formal methods are proposed, such as Timed Automata~\cite{bibitem6} and UPPAAL~\cite{bibitem7}, which can create, simulate and verify the timed automata using model checking techniques. Another example is Statecharts~\cite{bibitem5}, which is a modeling environment for state machine design. The BIP model is also used for component-based software modeling and system validation \cite{bibitem19}\cite{bibitem20}. Though these tools are well-done in modeling and verification, they are not professional in particular fields.

\begin{figure}[H]
	\centering 
	\includegraphics[scale=0.31]{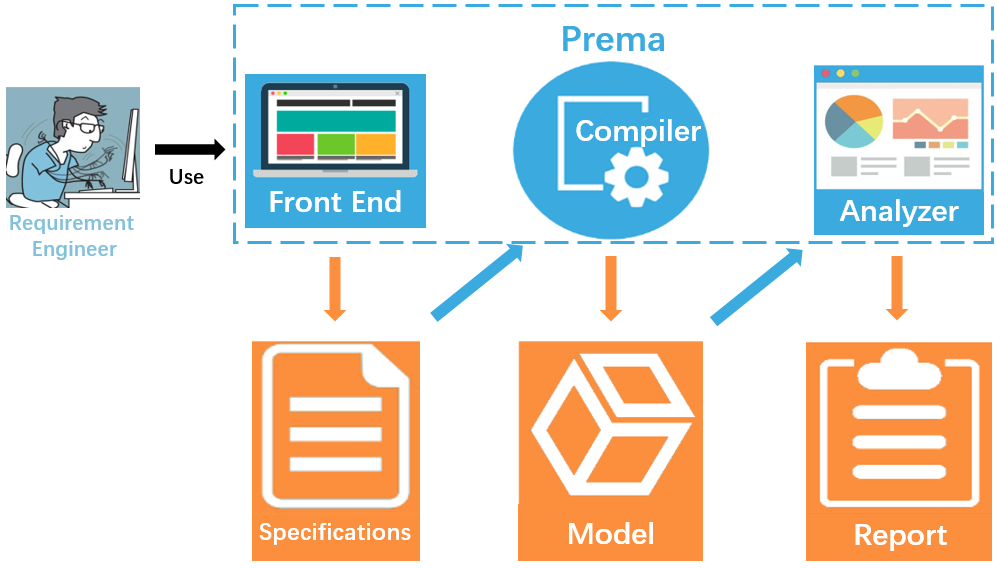}
	\caption{The framework of \emph{Prema}}
	\label{fig:FrameworkOfMethod}
\end{figure}

\par What's more, formal methods are not widely adopted in industry, even for safety-critical systems~\cite{bibitem3}. One major reason is that most software requirement engineers are not familiar with formal methods. Moreover, based on our collaborating experience with our industry partners, requirement engineers have little motivation to learn these methods due to the learning costs and the lack of requirement tools.

% For example, model checking works on formal models which may be generated from formal specifications. 

\par To address this problem, we present~\emph{Prema}, an requirements engineering tool that (1) uses our designed formal specification language for the industry partners; (2) allows users to write formal specifications and supplement natural language ones as comments in the editor, and (3) automatically generates rigorous V\&V reports based on the formal specifications. To reduce requirement engineers' learning costs, we custom-make the formal language based on the requirement writing habits to support the domain features.

\par The intuition for designing these customized formal languages came from the requirement documents of the three various industry fields: aerospace, aviation, and railway signal control. In each field, we choose a classical company to collaborate with. In railway signal control area, CASCO~\cite{bibitem18} is our partner. We found that in embedded control systems, most natural language specifications are very detailed and have similar styles. With some alterations, they can become formal specifications. For example, in CASCO's requirements documents, many specifications are written with pieces of Python-like pseudocode to clarify the corresponding specifications. With little modification, we designed a formal language just like the pseudocode, on which we then run the analysis. Besides CASCO, we also have designed two formal languages for the aerospace company and the aviation company respectively. These languages have been applied successfully in the safety-critical area in our previous work~\cite{bibitem16}~\cite{bibitem17}. As a proof of a concept, we present in this paper the demo of \emph{Prema} with the formal language for CASCO.

\par Our tool, \emph{Prema}, is an integrated development environment for requirements engineering, which consists of a front-end with an editor, a compiler, and an analyzer (shown in Figure~\ref{fig:FrameworkOfMethod}). The programming language of \emph{Prema} is C\# and it is running on Windows 10. Requirement engineers can write natural language and formal specifications, add images, tables, and flow charts in the editor. The compiler processes formal specifications into models. The analyzer contains a set of V\&V methods including (1) diagrams used for validation, such as state machine diagram; (2) system simulation; (3) test case generation for each requirement; and (4) property verification. In \emph{Prema}, natural language specifications and formal ones are developed and maintained in the same editing zone so that V\&V methods can be executed as soon as possible and requirement engineers do not need to switch environments for maintaining these two types of requirements artifacts.

\begin{figure}[H]
	\centering 
	\includegraphics[scale=0.42]{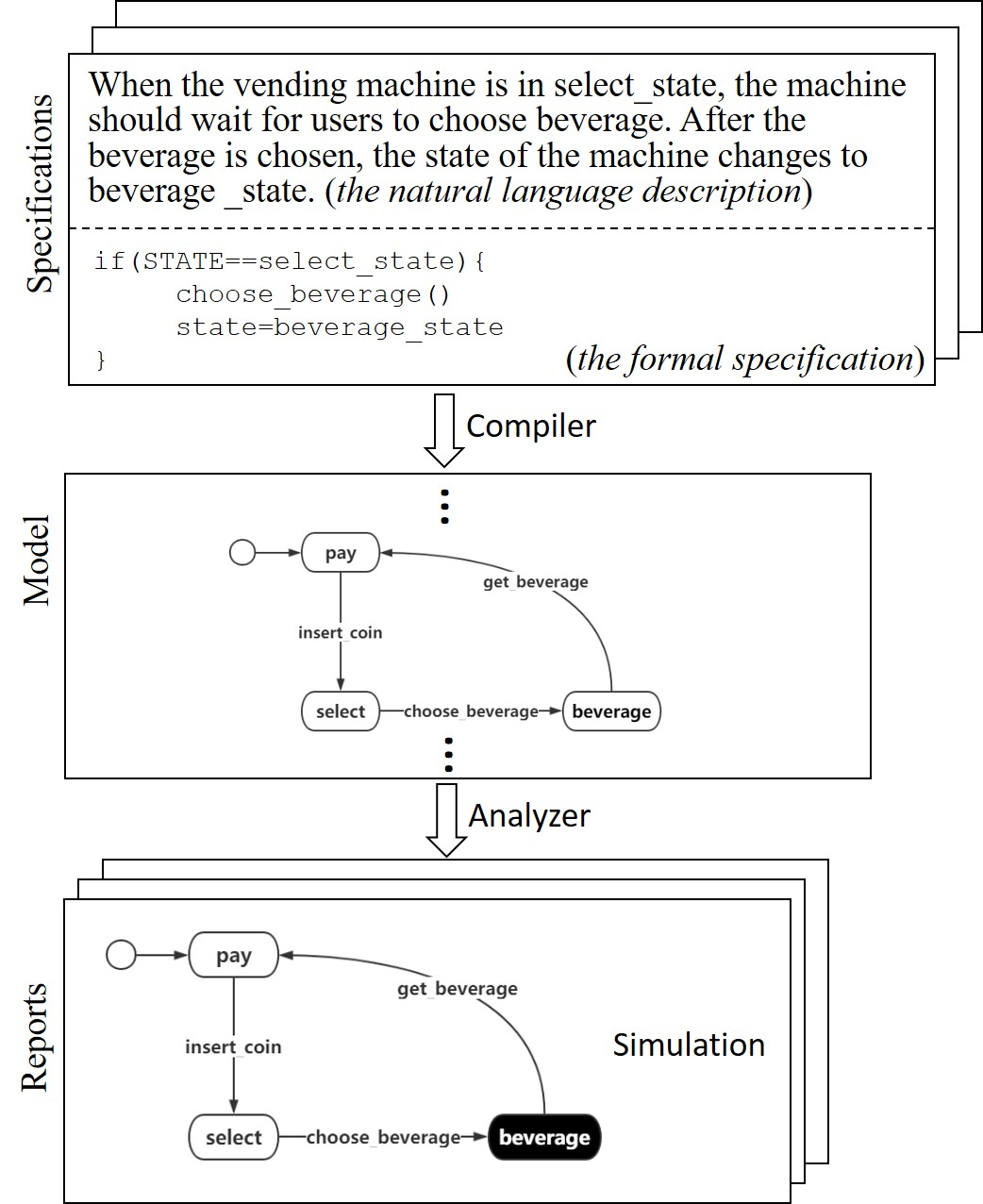}
	\caption{Vending Machine Example}
	\label{fig:VendingMachineExample}
\end{figure} 

\par We conducted a case study using \emph{Prema} for ATP requirements of CASCO. The requirements contain 454 pages, 482 computation tasks, 2 state machines, 1227 variable relationship diagrams. \emph{Prema} found 132 errors of the ATP requirements by using syntax analysis and V\&V methods. The errors include 127 syntax errors, 3 variable circular definition error, 1 variable dimension error and 1 DivideByZeroException.

\section{RUNNING EXAMPLE}

\par We use a simple application example of a vending machine that sells beverage, which is a classical V\&V example~\cite{bibitem9}. This vending machine has three states: pay\_state, select\_state, and beverage\_state. In the pay\_state, the system waits for users to insert coins. In the select\_state, the system waits for users to select beverage. In the beverage\_state, the system dispenses the beverage and transits to the pay\_state. 

\par Firstly, requirement engineers write specifications of this vending machine in \emph{Prema}. An example of a requirement is in Figure~\ref{fig:VendingMachineExample}, which has two parts: a natural language description of what this vending machine should do and a formal specification of the same requirement. After the specifications are written, requirement engineers run the compiler of \emph{Prema} which converts the formal specifications into a model saved in the backend (Model in Figure~\ref{fig:VendingMachineExample}). Finally, requirement engineers can run any V\&V analysis method provided in \emph{Prema}. Figure~\ref{fig:VendingMachineExample} shows the simulation of the state machine based on the specifications of the vending machine.

\section{APPROACH}

\par As shown in Figure~\ref{fig:FrameworkOfMethod}, front end, compiler and analyzer are the three parts of \emph{Prema}. Specifications, models and reports are the corresponding outputs of these three parts.

\begin{figure}[H]
	\centering 
	\includegraphics[scale=0.30]{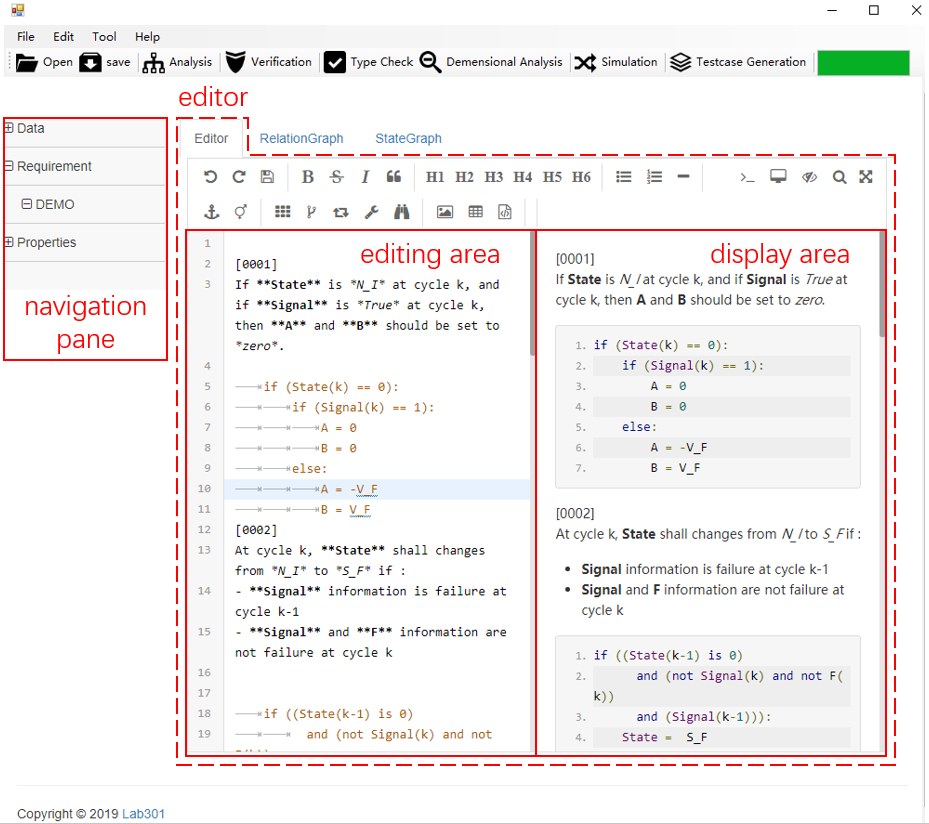}
	\caption{The Front End of \emph{Prema}}
	\label{fig:FrontEnd}
\end{figure} 

\subsection{Front-end}
\par A snapshot of the front-end is shown in Figure~\ref{fig:FrontEnd}. On the top of the front-end, there are six buttons which correspond to the six functions: analysis, verification, type check, dimensional analysis, simulation and test case generation. These are all V\&V methods which will be described in Section~\ref{sec:analyzer}.

\par The navigation pane shows the section headings of the requirements hierarchically. To open a requirement, users can click the heading of a requirement and the specifications of the requirement will be presented on the editor.

\par The editor supports Markdown syntax. By using the Markdown language, requirement engineers can easily add descriptions in formats other than text, such as truth tables and flow charts. Editing area is for writing requirements while the display area shows the live preview of the document. To distinguish the natural language specifications and formal ones in the editor, we just take the text segments with special symbols (TAB character) at the beginning into account as formal specifications, while others are natural language ones. The entire front-end is similar to an IDE, so that requirement engineers can use it without much learning cost.

\subsection{Compiler}

\par \emph{Prema} uses different compilers for different formal languages. A compiler has two parts: a parser and a model generator. The parser is produced by a third-party tool from a grammar defined by us. It can convert formal specifications into a syntax tree in accordance to the above-mentioned grammar. Then, a model generator will traverse the syntax tree, get all the information of it and transform it into a model.

\par
\begin{figure}[H]
	\centering 
	\includegraphics[scale=0.4]{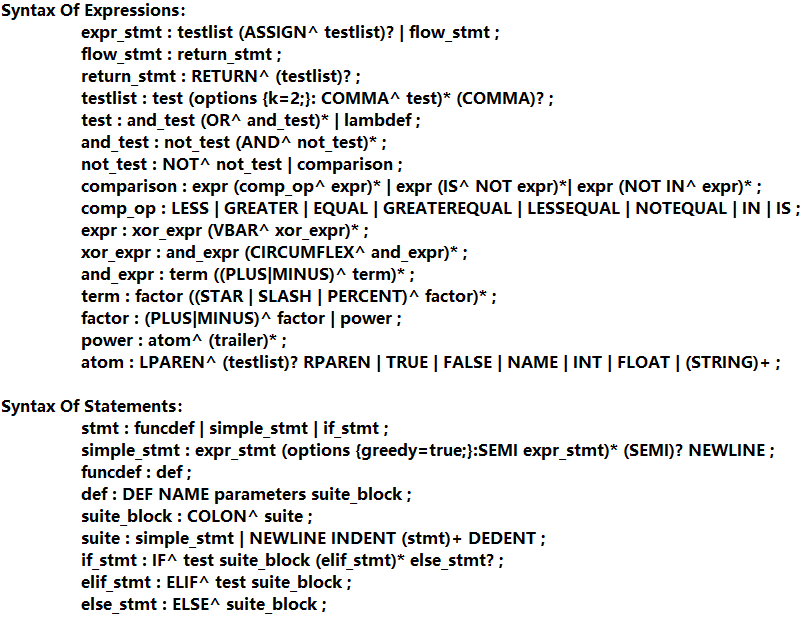}
	\caption{Syntax of the Formal Language for CASCO}
	\label{fig:Syntax}
\end{figure}

 The third-party tool we used to build the parser is ANTLR (ANother Tool for Language Recognition)~\cite{bibitem1}, which is a parser generator for processing structured text. Because our industry partners have different applications with various requirement writing habits and standards, the formal languages are customized to each company and have their own distinctive grammars. Although the formal language for CASCO's requirements shown in this paper has a grammar similar to that of Python's, our other formal languages are more like C++ programming language for the aviation company and the aerospace company. We list the grammar for the formal language we designed for CASCO in Figure~\ref{fig:Syntax}.

\par After we get a syntax tree from a parser, the model generator creates a model based on the tree. To describe an embedded control system accurately, the model usually contains three parts: a state machine, computation tasks and a data dictionary. The state machine shows the states of the system and the migration relationships between them. The model generator automatically labels the variables that have names ended with "State" as state variables. The data dictionary records all the variables used in the requirements. The validation of the model is proved to be equivalent to that of the original formal requirements. We present the model we created for a subsystem in the railway signal control system of CASCO in our case study (Section~\ref{sec:casestudy}). The model is a little different from the universal model of embedded control software. It has two small state machines which represent part of the system. The other two formal languages (for the aerospace company and the aviation company) are transformed into different models. For example, the model for aviation company has control flow graphs, in addition to a state machine and a data dictionary.

\subsection{Analyzer}
\label{sec:analyzer}

\par After we have the model generated by the compiler, the analyzer helps to check whether the model is consistent with the expectations of the requirement engineers for software functionalities. The analyzer has the following parts: (1) diagram-based validation, (2) simulation, (3) test case generation, (4) property verification, (5) type check, and (6) dimensional analysis. Each part has a corresponding button on the front-end, so requirement engineers can click a button to run that V\&V method. Due to the space limit, we will not discuss the last two parts: type check and dimensional analysis.

% \begin{figure}[tb]
% 	\centering 
%     \includegraphics[scale=0.4]{SchemaMigrationDiagram.jpg}
%     \caption{mode transition Diagram (4 schemas, 6 migration conditions)}
%     \label{fig:SchemaMigrationDiagram}
% \end{figure} 

\subsubsection{Diagrams for Validation}
\par \emph{Prema} can generate mode transition diagrams, state machine diagrams and variable relationship diagrams. A mode transition diagram describes the operating framework of the embedded control software, which is represented as an automata. An example of a state machine diagram is shown in Figure~\ref{fig:VendingMachineExample}. In the state machine diagram, it shows the relationship between multiple states of a variable. The states are shown as vertices while the transition relationships are represented by the directed edges with conditions on them. In the variable relationship diagram, users can select the variable which needs to be checked in the list supported by \emph{Prema}. This variable is called key variable. The diagram will show the variables which use the key variable and the variables used by the key variable. \emph{Prema} supports direct and indirect reference relationship among the variables.

% \begin{figure}[tb]
% 	\centering 
%     \includegraphics[scale=0.4]{StateMachineDiagram.jpg}
%     \caption{State Machine Diagram}
%     \label{fig:StateMachineDiagram}
% \end{figure} 

\subsubsection{Simulation}
\par
\begin{figure}[H]
	\centering 
	\includegraphics[scale=0.41]{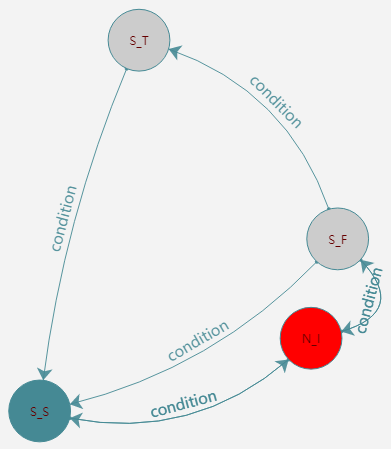}
	\caption{Simulation Sketch Map}
	\label{fig:simulation}
\end{figure}

 The simulation function allows users to run a simulation of the system behavior of software at the requirement level. It can be used for all computation tasks of the state machine.  When using this function, users have to choose some requirement items which need to be simulated from the entire requirements. Figure~\ref{fig:simulation} shows the simulation of a state machine. This state machine has four states and seven migration conditions. There is one migration condition on each directed edge. To run a simulation, the users need to provide input variable values at each cycle or give a list of input variable values of many cycles at once. In addition, because it is a simulation of periodic software, so in each cycle, users can stop the simulation to judge whether the results are expected. 

\subsubsection{Test case Generation}
\par The test case generation function can create test cases for each requirement. The generation algorithm is based on MC/DC (Modified Condition Decision Coverage)~\cite{bibitem10} criteria, which is required by DO-333~(a industrial standard)~\cite{bibitem15}. According to the definition of MC/DC, each condition in each decision must be able to independently affect the output of a decision. So we designed an algorithm to collect constraints of each function. Through using the Z3 solver~\cite{bibitem8}, we can get test cases by constraints.

\subsubsection{Property Verification}
\par The property verification part is where we apply formal methods to check the safety of the requirements. These methods can help requirement engineers discover unexpected behaviors expressed in the requirement. To use this function, engineers first provides a property which is expected to be satisfied by the system in a Boolean expression. Then they selects a part of the requirements that need to be verified. \emph{Prema} uses Z3 solver to solve the expression. If Z3 does not output counterexample, it means the property is satisfied. Otherwise error paths and variable values which lead to failure will be displayed.

\section{CASE STUDY}
\label{sec:casestudy}

\par The case study examines how \emph{Prema} works for an entire requirement document of a subsystem in a railway signal control system of CASCO.  The original requirements document of the ATP system is written in Microsoft Word, which has 454 pages. It contains 482 computation tasks, which can be viewed as functions or programming blocks. The requirements not only have English and Chinese description of requirements, but also have images and code snippets. Because the requirement documents provided by CASCO are Word documents, so we use the import function of \emph{Prema} to transform it into Markdown documents, which are shown in the front end. 
\par The diagram-based validation function generates 2 state machine diagrams with 197 variables and 1227 variable relationship diagrams. The two state machines both have 4 vertices which represent the states of the state variable with 7 and 8 directed edges in them respectively. Through using the diagram-based validation function, we detect 3 variable circular definition error of the computation tasks.
\par Simulation is executed on the 2 state machines with input variable values of 15273 cycles. In the simulation process, there exists a DivideByZeroException which interrupts the program and reports an error.
\par Property verification is used on the entire requirement with the property supported by the requirement engineers. The result is consistent with the expectation of them. 
\par Test case generation is based on MC/DC criteria. In ATP, 131 computation tasks can generate test cases. Each of them has 6.8 test cases on average. According to statistics, the requirement items whose test case coverage is 100\% account for 64.85\% of the entire requirements. 

\section{CONCLUSION}
\label{sec:conclusion}
\par In this paper, we presented \emph{Prema}, a tool for precise requirement editing, modeling and analysis. The case study of ATP shows how it helps requirement engineers to efficiently complete their work and finding potential requirement errors.

\section{ACKNOWLEDGEMENTS}
\label{sec:acknowledgements}
\par We thank the anonymous reviewers for their valuable
feedback. Yihao Huang is partially supported by NSFC Projects No. 61572197 and No. 61632005. Weikai Miao is supported by the NSFCs of China (No. 61872144 and No. 61872146 and No. 61532019). Geguang Pu is partially supported by China HGJ Project No. 2017ZX01038102-002 and NSFC Project No. 61532019. This work is also supported by the CASCO Ltd..

\bibliographystyle{IEEEtran}

	% \bibliography{reference}
\end{document}